\documentstyle[epsf,multicol,prl,aps]{revtex}
\newcommand{\rmP}{{\rm p}}
\newcommand{\rmt}{{\rm t}}
\newcommand{\rmd}{{\rm d}}
\newcommand{\rmR}{{\rm R}}
\newcommand{\rmA}{{\rm A}}
\newcommand{\rmRS}{{\rm RS}}
\newcommand{\Det}{{\rm Det}}
\newcommand{\Tr}{{\rm Tr}}
\newcommand{\calH}{{\cal H}}
\newcommand{\calC}{{\cal C}}
\newcommand{\calP}{{\cal P}}
\newcommand{\calT}{{\cal T}}
\newcommand{\calL}{{\cal L}}
\newcommand{\calZ}{{\cal Z}}
\newcommand{\calD}{{\cal D}}
\newcommand{\bmC}{ \mbox{\boldmath $C$} }
\newcommand{\tr}{{\rm tr}}
\newcommand{\bra}[1]{\langle #1|}
\newcommand{\ket}[1]{|#1\rangle}
\newcommand{\tpsi}{{\widetilde\psi}}
\newcommand{\otimesshort}{{\!\!\otimes\!\!}}
\begin{document}
\draft
\preprint{}
\title{Disordered $d$-wave superconductors with chiral symmetry} 
\author{ Takahiro Fukui\cite{Email}} 
\address{Institut f\"ur Theoretische Physik,
Universit\"at zu K\"oln, Z\"ulpicher Strasse 77,
50937 K\"oln, Germany}
\date{May 26, 1999}
\maketitle
\begin{abstract}
A two-dimensional lattice model for $d$-wave superconductor 
with chiral symmetry is studied.
The field theory at the band center is shown to be
in the universality class of U($2n$)/O($2n$) and
U($2n$) nonlinear sigma model for the system with broken and
unbroken time-reversal symmetry, respectively.
Vanishing of the beta function implies extended states at the 
band center.
Density of state vanishes 
as a cubic function of the energy at the band center 
for the former case, while 
linear for the latter.
\end{abstract}

\begin{multicols}{2}

Symmetries play a crucial role in random critical 
phenomena. 
The rotational and time-reversal invariances specify
the well-known universality classes whose underlying
symmetries are orthogonal, symplectic, and unitary \cite{UniCla}.
The variety of universality classes, however, has
turned out to be richer.
Above all,
Altland and Zirnbauer \cite{AltZir,Zir} have recently discussed  
universality classes for quasi-particles
in disordered spin-singlet BCS Hamiltonian, 
denoted in their article by $C$ and $C$I 
for the case with
broken and unbroken time-reversal symmetry, respectively. 
These universality classes are actually involved with  
vortices in $s$-wave superconductors \cite{BCSZs},
quasi-particle transport in $d$-wave superconductors
\cite{SFBN,BCSZd},
and the spin quantum Hall (QH) transitions \cite{SQH}.
Especially, dirty $d$-wave superconductors have attracted a lot of
interest, because
they can be described by Dirac Fermions \cite{NTW}
due to gapless quasi-particle spectrum.
Random Dirac Fermions actually give 
nontrivial critical points \cite{LFSG}.
The criticality of the density of state (DOS) 
of disordered $d$-wave superconductors is, for example,
still controversial \cite{NTW,ZHH,SFBN}.

Another possibility is the class for 
the two-sublattice model \cite{Gad},
where Hamiltonian has a special symmetry which will be specified
momentarily.
This symmetry will be referred to as chiral symmetry \cite{ChiRan}.
One example in two-dimension is the random flux model \cite{RFM}
and another is the random hopping 
fermions with $\pi$-flux \cite{RHM}, 
both of which have an isolated delocalized states at the band center.
Various universality classes, including those mentioned above,
are well summarized by Zirnbauer \cite{Zir}.

In this paper, we study a system with chiral 
symmetry in $d$-wave superconductors.
The basic idea is that 
the pure $d$-wave lattice system is 
a kind of two-sublattice model.
It is, therefore, natural to include randomness keeping the 
same symmetry of the pure Hamiltonian.
The model we study has 
a SU(2) symmetry as well as the chiral symmetry. 
It is also interesting to study a system 
with such enhanced symmetries. 
We show that the field theory of the model with broken and
unbroken time-reversal symmetry 
is in the universality class of U($2n$)/O($2n$) and
U($2n$) nonlinear sigma model (NLSM), respectively.
It is interesting to compare the classes to those of
two-sublattice models without SU(2) symmetry, which are
in the class of U($n$) and U($2n$)/Sp($n$) NLSM
or to those of $d$-wave superconductors without chiral symmetry,
which are in the class of Sp($n$)/U($2n$) and Sp($n$)
NLSM for the case with
broken and unbroken time-reversal symmetry, respectively.
We conclude that regardless of the time-reversal symmetry
the present model has extended states at the band-center, 
which are associated with 
diffusive spin transport \cite{SFBN,BCSZd}.
We discuss the behavior of the quasi-particle density of state (DOS),
which should vanish at the band center as $E^3$ and $E$ 
in broken and unbroken time-reversal cases,
respectively.

Let us start from the lattice 
Hamiltonian of a $d$-wave superconductor
defined on the square lattice in two-dimension \cite{SFBN,BCSZd},
\begin{equation}
H=\sum_{\langle i,j\rangle}
\left(
  t_{ij}\sum_{\sigma}c_{i\sigma}^\dagger c_{j\sigma}
 +\Delta_{ij}  c_{i\uparrow}^\dagger   c_{j\downarrow}^\dagger
 +\Delta_{ij}^*c_{i\downarrow}         c_{j\uparrow}
\right).
\label{LatHam}
\end{equation}
Hermiticity and SU(2) symmetry (spin-rotation) require 
$t_{ij}=t_{ji}^*$ and $\Delta_{ij}=\Delta_{ji}$, respectively.
In the absence of randomness,
the pure $d$-wave Hamiltonian is given 
by $t_{j,j\pm\hat x}=t_{j,j\pm\hat y}=-t_0$,
$\Delta_{j,j\pm\hat x}=-\Delta_{j,j\pm\hat y}=\Delta_0$, 
and others are zero,
where $\hat x=(1,0)$ and $\hat y=(0,1)$.
It is easy to see that the pure Hamiltonian change the sign under the 
transformation 
$c_{j\sigma}\rightarrow (-)^{j_x+j_y}c_{j\sigma}$,
which is the chiral symmetry as mentioned-above.
Therefore, the pure system has the SU(2), time-reversal, and 
chiral symmetry.
By introducing disorder which breaks some symmetries explicitly
but keeps the others, the model shows various kinds of 
universality classes.
Those studied so far are as follows:
If we keep only SU(2) symmetry, the (replicated) model
has Sp($n$) symmetry, which is spontaneously
broken to U($2n$). 
If we keep in addition 
the time-reversal invariance, the model has an enhanced symmetry 
Sp($n$)$\times$Sp($n$), which is broken to Sp($n$) 
\cite{SFBN,BCSZd}.
In this paper, the symbol $\langle i,j\rangle$
in Eq. (\ref{LatHam}) is restricted to the nearest neighbor
pairs, which keeps the chiral symmetry,
and universality classes of the $d$-wave superconductors
unknown so far are investigated.

The pure Hamiltonian has four nodes, where gapless quasi-particle
excitations exist \cite{NTW}. 
To study the low-energy properties of the system 
governed by them, let us take 
the continuum limit near the nodes \cite{SFBN}, 
\begin{eqnarray}
c_{j\uparrow}/a\sim&&
  i^{ j_x+j_y}\chi_{\uparrow11}(x)
 -i^{-j_x-j_y}\chi_{\downarrow21}(x) 
\nonumber\\&& 
 +i^{-j_x+j_y}\chi_{\uparrow12}(x)
 -i^{ j_x-j_y}\chi_{\downarrow22}(x) ,
\nonumber\\
c_{j\downarrow}/a\sim&&
  i^{ j_x+j_y}\chi_{\downarrow11}^\dagger(x)
 +i^{-j_x-j_y}\chi_{\uparrow21}^\dagger(x)  
\nonumber\\&& 
 +i^{-j_x+j_y}\chi_{\downarrow12}^\dagger(x)
 +i^{ j_x-j_y}\chi_{\uparrow22}^\dagger(x) ,
\end{eqnarray}
where $a$ is a lattice constant, $x=aj$, and
indices $\sigma, a,$ and $i$ of the field $\chi_{\sigma ai}$ are 
associated with spin, particle-hole, and nodes, respectively.
Namely, the field $\chi$ at each $x$ lives in the space
$V=\bmC^2\otimes \bmC^2\otimes \bmC^2$.
The pure Hamiltonian \cite{NTW,SFBN} is then 
$
H_\rmP=
\int d^2x\chi^\dagger\calH_\rmP\chi,
$
where
\begin{equation}
\calH_\rmP=
1_2\otimes
\left(
 \begin{array}{ll}
  -v\gamma_\mu i\partial_\mu & \\
   &   (x\leftrightarrow y)
 \end{array}
\right),
\label{ConPurHam}
\end{equation}
and the coordinates have been transformed as 
$x,y\rightarrow\frac{\pm x+ y}{\sqrt{2}}$.
Here the explicit matrix in Eq. (\ref{ConPurHam}) denotes 
the node space, and 
gamma matrices denotes the particle-hole space, defined by 
$\gamma_1= r\sigma_3$ and $\gamma_2=r^{-1}\sigma_1$ 
with $r=\sqrt{v_F/v_\Delta}$, and $v=\sqrt{v_Fv_\Delta}$,
where $v_F=2\sqrt{2}ta$ and $v_\Delta=2\sqrt{2}\Delta_0$.

The symmetries of the lattice model are translated into the 
continuum model as
\begin{equation}
 \begin{array}{ll}
  \calC\calH\calC^{-1}=-\calH^\rmt , \qquad   &
  \calC=i\sigma_2\otimes1_2\otimes1_2,        \\
  \calP\calH\calP^{-1}=-\calH,                &      
  \calP=i\sigma_2\otimes i\sigma_2\otimes1_2, \\ 
  \calT\calH\calT^{-1}=-\calH ,               &
  \calT=\calC\calP ,
\end{array}
\label{Sym}
\end{equation}
which describe, respectively, 
the SU(2), chiral, and time-reversal symmetry.
The total Hamiltonian density
is given by $\calH=\calH_\rmP+\calH_\rmd$,
where $\calH_\rmd$ is disorder 
potentials satisfying Eq. (\ref{Sym}).
In what follows, we study the model with broken 
and unbroken time-reversal symmetry at the same time.
For the former case, the third condition should be omitted.
It is easy to write down explicit disorder potentials
using these conditions, though it is not necessary
in the following calculation.

To study the DOS and the conductance of the spin transport 
\cite{NLSM},
let us define the Green functions,
\begin{eqnarray}
&&
G(x)=\tr_V
 \bra{x}(i\epsilon-\calH)^{-1}\ket{x},
\nonumber\\
&&
K(x,x')=\tr_V
 \left|
  \bra{x} (i\epsilon-\calH)^{-1}\ket{x'}
 \right|^2 ,
\end{eqnarray}
where $\tr_V$ is the trace in the $V$ space.
Although it may be easy to introduce the generating functional
of these Green functions, we have to prepare some notations.
Firstly, the generating functional 
expressed by path-integrals over Fermi fields is
defined in a standard way by introducing replica, 
$\chi_{i}\rightarrow\chi_{i\alpha}$ and 
$\bar\chi_{i}\rightarrow\bar\chi_{\alpha i}$,
where $i$ and $\alpha$ are indices denoting $V$ and 
replica space $W_\rmR=\bmC^n$, respectively.
Lagrangian density is then
$\calL=-\tr_{W_\rmR}\bar\chi\left(i\epsilon-\calH\right)\chi$,
where $\tr_{W_\rmR}$ is the trace in the replica space.
Moreover, we introduce an auxiliary space \cite{Zir}
to reflect the symmetries
in the $V$ space to an auxiliary field introduced later 
[See Eq. (\ref{SymQ})],
$W_\rmR\rightarrow W=W_\rmR\otimes W_\rmA$, with
$W_\rmA=\bmC^2\otimes\bmC^2\otimes\bmC^2$,
where each space in $W_\rmA$ is associated with the 
SU(2), chiral, and time-reversal symmetry, respectively,
and in this extended space, Fermi fields are denoted by
$\tpsi_{\alpha i}$ and $\psi_{i\alpha}$, which are subject to 
\cite{Zir},
\begin{equation}
\begin{array}{ll}
 \tpsi=  \gamma \psi^\rmt \calC^{-1}, \qquad
&\psi =  \calC \tpsi^\rmt \gamma^{-1},\\
 \tpsi= i\tau  \tpsi      \calP^{-1}, 
&\psi = i\calP  \psi      \tau^{-1},  \\
 \tpsi= i\pi   \tpsi      \calT^{-1}, 
&\psi = i\calT  \psi      \pi^{-1} .
\end{array}
\label{AuxCon}
\end{equation}
Matrices $\gamma$, $\tau$, and $\pi$ are defined in the $W$ space,
given by
$\gamma=1_n\otimes i\sigma_2\otimes 1_2      \otimes 1_2     $,
$\tau  =1_n\otimes 1_2      \otimes i\sigma_2\otimes 1_2     $, and
$\pi   =1_n\otimes 1_2      \otimes 1_2      \otimes \sigma_1$.
The identity
$\tr_W(\omega\tpsi\psi-\tpsi\calH\psi)
 =\tr_{W_\rmR}\bar\chi(i\epsilon-\calH)\chi$,
where 
$\omega=i\epsilon 1_n\otimes\sigma_3\otimes\sigma_3\otimes\sigma_3$,
leads to the generating functional
$\calZ=\int\!\!\calD\bar\psi\calD\psi e^{-S}$ with
\begin{equation}
S=-\int\!\!d^2x\tr_W
 \left(
  \omega\tpsi\psi-\tpsi\calH\psi+J\tpsi\psi
 \right).
\end{equation}
Green functions are expressed as
\begin{eqnarray}
&&
G(x)=d_{W_\rmA}\lim_{n\rightarrow0}
\left\langle
 (\tpsi\psi)_{11}(x)
\right\rangle,
\nonumber\\
&&
K(x,x')=-d_{W_\rmA}^2\lim_{n\rightarrow0}
\left\langle
 (\tpsi\psi)_{12}(x')(\tpsi\psi)_{21}(x)
\right\rangle,
\label{GreFun}
\end{eqnarray}
where 
$1= \alpha \otimes\!\!\uparrow
\otimesshort\uparrow\otimesshort\uparrow$ and
$2= \beta \otimes\!\!\downarrow
\otimesshort\uparrow\otimesshort\uparrow$, 
and $d_{W_\rmA}$ is the dimension of the auxiliary space,
given by 4 and 8 for 
broken and unbroken time-reversal case, respectively.
Assume that disorder potentials obey the Gaussian 
distribution
$\int\calD\calH_\rmd\exp(-\frac{1}{2g}\int d^2x\tr_V\calH_\rmd^2)$.
Then it is easy to average over them, as usual.
Here, some comments may be useful \cite{Zir}:
Firstly, disorder potentials are integrated out by using
$-\frac{1}{2g}\tr_V\calH_\rmd^2+\tr_V\calH_\rmd\psi\tpsi
=-\frac{1}{2g}(\tr_V\calH_\rmd-g\psi\tpsi)^2
+\frac{g}{2}\tr_V(\psi\tpsi)^2$.
It turns out that the integration over $\calH_\rmd$ is automatic
because $\psi\tpsi$ satisfy the same symmetries 
as those of $\calH_\rmd$ due to Eq. (\ref{AuxCon}). 
Secondly, $\tr_V(\psi\tpsi)^2=-\tr_W(\tpsi\psi)^2$, which is
converted into Yukawa-type interactions by introducing
an auxiliary matrix field defined in the $W$ space into the action
in the form
$\frac{1}{2g}\tr_W (Q+g\tpsi\psi-\omega)^2$.
Integrating out the Fermi fields, we end up with an effective action, 
\begin{eqnarray}
S=&&-\frac{1}{2g}\int\!\!d^2x\tr_W
\left(
Q^2-2Q\omega
\right)
\nonumber\\&& 
-\frac{1}{d_{W_\rmA}}\ln\Det_{V\otimes W}
\left(
1\otimes Q-\calH_\rmP\otimes1
\right) .
\label{EffAct}
\end{eqnarray}
Here we have set $J=0$ for simplicity and
the anti-Hermitian
auxiliary field $Q=-Q^\dagger$ is subject to
\begin{equation}
Q=-\gamma Q^\rmt\gamma^{-1},\quad
Q=-\tau   Q     \tau^{-1},  \quad
Q=-\pi    Q     \pi^{-1} .  
\label{SymQ}
\end{equation}

To get a saddle-point solution, set $\omega=0$,
and assume $Q$ is diagonal and spatially constant.
Then, saddle-point equation reduces to 
\begin{equation}
\frac{1}{g}Q_0=
-\frac{1}{d_{W_\rmA}}\tr_V(Q_0-\calH_\rmP)^{-1}(x,x),
\end{equation}
which gives 
$Q_0=iv_01_n\otimes\sigma_3\otimes\sigma_3\otimes\sigma_3$
as a solution. Here 
$v_0\sim\Lambda\exp(-\pi d_{W_\rmA}v^2/4g)$ with $\Lambda$ being
high-energy cut-off.
This solution gives rise to an exponentially small density of state
at the band-center.

The effective action has U($2n$) and U($2n$)$\times$U($2n$)
symmetry for the model with 
broken and unbroken time-reversal symmetry,
respectively.
To be concrete, let us consider the former case.
For the action function to be invariant under the group action
$Q\rightarrow gQg^{-1}$, $g$ should satisfy
$\gamma=g\gamma g^\rmt$ and $\tau g\tau^{-1}=g$.
They are explicitly given by
\begin{equation}
g=V
\left(
 \begin{array}{ll}
  u & \\ & \overline{u}
 \end{array}
\right)V^\dagger,
\quad
Q=V
\left(
 \begin{array}{ll}
   & q\\ -q^\dagger&
 \end{array}
\right)V^\dagger,
\label{ExpGQ}
\end{equation}
where matrices in Eq. (\ref{ExpGQ})
denote the chiral space of $W$,
and $u$ is $2n\times2n$ unitary matrix in the replica and 
SU(2) space of $W$, while $q$ is a complex matrix
with a condition $q^\rmt=q$.
$V$ is a matrix defined by $V=V_1V_2$ with
$V_1=1_n\otimes \mu\otimes \mu$ and
$V_2=1_n\otimes 1_2    \otimes\frac{1}{2}(1_2+\sigma_3)
    +1_n\otimes\sigma_2\otimes\frac{1}{2}(1_2-\sigma_3)$,
where $\mu=\frac{1}{\sqrt{2}}(\sigma_3+\sigma_2)$.
It is, therefore, easy to show that the saddle-point solution
is invariant under the group action if $u^\rmt u=1$.
Thus, the U($2n$) symmetry is spontaneously broken 
to O($2n$), and the saddle-point manifold is
U($2n$)/O($2n$).

Local quantum fluctuation 
around the saddle-point
yields Goldstone mode, which governs the low energy properties
of the present system.
To derive an effective 
action of this mode integrating out the massive mode,
let us decompose the field $Q$ 
into transverse and longitudinal modes as 
$Q(x)=T(x)(Q_0+L(x))T^\dagger(x)$, where
\begin{equation}
T=V
 \left(
 \begin{array}{ll}
  U & \\ & \overline{U}
 \end{array}
\right) 
  V^\dagger, \quad
L=V
 \left(
 \begin{array}{ll}
    &P \\ -P& 
 \end{array}
\right) 
  V^\dagger .
\end{equation}
Here
$U(x)\in$ U($2n$)/O($2n$) and
$P(x)$ is a real matrix with $P^\rmt=P$.
Next transformation properties of $U(x)$ and $P(x)$ fields
should be examined.
The action of global $u\in$ U($2n$) induces
$uU(x)=U'(x)h(u,U(x))$, where $U'(x)\in$ U($2n$) and
$h(u,U)\in$ O($2n$). It should be noted that $h(u,U)$ is a 
nonlinear function of $U(x)$ dependent on $x$.
Therefore, we have the following transformation laws of 
$U$ and $P$,
$U\rightarrow uUh^{-1}$ and
$P\rightarrow hPh^{-1}$.

Keeping this in mind, we proceed to calculate the action of
the Goldstone mode.
The point is the derivative expansion of the $\ln\Det$ term
in Eq. (\ref{EffAct}),
\begin{eqnarray}
&&\ln\Det_{V\otimes W}(1\otimes Q-\calH_\rmP\otimes1)
\nonumber\\&& 
\sim
\Tr_{V\otimes W}\Delta_FG-\frac{1}{2}\Tr_{V\otimes W}
\left(\Delta_FG\right)^2 ,
\end{eqnarray}
where
$\Delta_F$ is the free propagator defined by
$\Delta_F^{-1}=1\otimes Q_0-\calH_\rmP\otimes1$ and $G$ is 
\begin{eqnarray}
G&&=(1\otimes T^\dagger)(\calH_\rmP\otimes1)(1\otimes T)
-\calH_\rmP\otimes1 
\nonumber\\
&&=
\left(
 \begin{array}{ll}
  v\gamma_\mu iT^\dagger\partial_\mu T &\\
   & (x\leftrightarrow y)
 \end{array}
\right).
\end{eqnarray}
The explicit matrix in the above denotes the node space in $V$.
It is stressed that the effective action
should be independent of the gauge associated with the local 
O($2n$) transformation.
To see this, let us write down the field $T^\dagger\partial_\mu T$
in $W$ space,
\begin{eqnarray}
T^\dagger\partial_\mu T=V
\left(
 \begin{array}{ll}
  V_\mu+ A_\mu &\\
  &V_\mu- A_\mu 
 \end{array}
\right)
V^\dagger,
\end{eqnarray}
where
$A_\mu, V_\mu=\frac{1}{2}
(U^\dagger\partial_\mu U\mp U^\rmt\partial_\mu\overline{U})$.
The transformation laws of these fields are
$A_\mu\rightarrow hA_\mu h^{-1}$ and
$V_\mu\rightarrow hV_\mu h^{-1}+h\partial_\mu h^{-1}$,
which tells that $V_\mu$ is a gauge field associated with
the hidden local O($2n$) symmetry \cite{BKY}.
In the leading order of the derivative expansion, 
therefore, only the gauge covariant $A_\mu$ 
field appears in the action.

By using these, manifestly gauge-invariant
action of the Goldstone mode can be obtained. 
The principal term is calculated as 
$\frac{1}{2d_{W_\rmA}}\Tr_{V\otimes W}(\Delta_FG)^2=
-\frac{2}{b}\Tr_{W_\rmRS}A_\mu^2$.
After gauge-fixing $U^\rmt =U$, we have
\begin{equation}
S=
\int\!\!d^2x\tr_{W_\rmRS}
\left[
  \frac{1}{2b}
   \partial_\mu U^2\partial_\mu U^{-2}
 -\frac{v_0\epsilon}{g}
  (U^2+U^{-2})
\right].
\label{ActGol}
\end{equation}
Here $W_\rmRS$ denotes a part of the space $W$ 
restricted to the replica and SU(2) spaces
omitting the chiral space,
and
$\frac{1}{b}=
\frac{1}{4\pi}\frac{v_F^2+v_\Delta^2}{v_Fv_\Delta}$.
This coupling constant is just the same as that derived by
Senthil {\it at al.} \cite{SFBN}.
So far we have obtained an effective action for 
broken time-reversal system.
It is easy to follow similar calculations for the unbroken 
time-reversal system.
The results are the same as Eq. (\ref{ActGol}) including
the coupling constant, but with 
$U(x)\in$ U($2n$) and
a breaking term
$\frac{2v_0\epsilon}{g}\tr_{W_\rmRS}
\Re(U^2+U^{-2})$.

In the process of the renormalization, it is shown that
the operator 
$ \frac{2}{c}\Tr^2_{W_\rmRS}A_\mu=
\frac{1}{2c}\Tr^2_{W_\rmRS}U^2\partial_\mu U^{-2}$ 
is needed.
In the leading order we have taken above, 
the coupling constant is $\frac{1}{c}=0$.
However, higher loop expansion of the $\ln\Det$ term actually gives 
a finite coupling constant.
After the replica limit $n\rightarrow0$,
the renormalization group equations at one-loop order
are given by \cite{Hik,Gad}
\begin{eqnarray}
&&
\frac{{\rm d}b}{{\rm d}l}=0, \quad
\frac{{\rm d}c}{{\rm d}l}=-\alpha_1 c^2,
\nonumber\\
&&
\frac{{\rm d}\epsilon}{{\rm d}l}=(2+\zeta)\epsilon,
\quad 
\zeta=\frac{1}{2}
\left(
 \beta_1 b+\frac{b^2}{c}
\right),
\end{eqnarray}
where $(\alpha_1,\beta_1)=(\frac{1}{4},\frac{1}{2})$ and
$(\frac{1}{2},0)$ 
for U($2n$)/O($2n$) and U($2n$) models, respectively.
The spin conductance is related only with the coupling constant 
$b$ as
$\frac{1}{b}=2\alpha_1\pi\sigma_{\rm s}$, which is obtained by
the bare diffusion constant calculated from Eq. (\ref{GreFun}).
Therefore, 
vanishing of the beta function for $b$ \cite{Com} implies 
{\it delocalized state at zero energy}, which is
due to the chiral symmetry.
On the other hand, the behavior of the DOS depends on the
coupling constant $c$.
Since $\epsilon$ has the same dimension
of the energy,
we can compute, according to Gade \cite{Gad}, 
a rough estimate of the DOS for finite $E$ as
$\rho(E)\sim \frac{\Lambda}{E}
e^{-4\sqrt{(\alpha_1b^2)^{-1}\ln(\Lambda/E)}}$.
This yields an enhancement of DOS near the zero energies.
However, the result from perturbative calculations
needs alternative consideration for DOS at the zero energy.

\begin{figure}[htb] 
\epsfxsize=60mm 
\centerline{\epsfbox{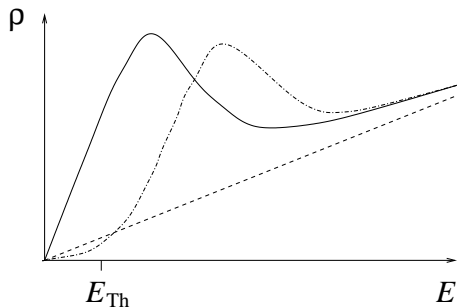}} 
\vspace{0.2cm}
\caption{Schematic illustration of the 
density of state as a function of energy.
Dashed-line denotes DOS of pure system, 
while dashed-dotted-line (full-line) 
denotes DOS for broken (unbroken) 
time-reversal symmetry.
The bump means the enhancement 
obtained by the renormalization group analysis.}
\label{f:dos}
\end{figure}\noindent

Since delocalization occurs at the zero energy, 
the localization length is quite long, i.e., the order of the 
system size $L$ near the zero energy. 
Accordingly, in the energy scale much smaller than the  
Thouless energy $\frac{D_{\rm s}}{L^2}$,
where $D_{\rm s}$ is a spin diffusion constant,
quasi-particles are diffusing all around the system keeping
the symmetry of the Hamiltonian, and hence,
the spatial dependence of the system is smeared out and 
we can describe the system by a random matrix theory.
Taking only zero mode of $Q$ into account,
we actually have from Eq. (\ref{EffAct})
an effective action of a random 
matrix theory with the symmetries (\ref{Sym}) but defined in 
$V=\bmC^{N}\otimes\bmC^2\otimes\bmC^2\otimes\bmC^2$, where $N=L^2$.
The Hamiltonian $\calH$ is now not a field but a
quantum mechanical one subject to Eq. (\ref{Sym}).
To calculate the DOS near the zero energy of such a
random matrix ensemble,
it may be convenient to rotate the basis by a
orthogonal transformation and 
to switch into more convenient basis.
We can choose 
$\calC=1_{N}\otimes1_2\otimes i\sigma_2\otimes1_2$,
$\calP=1_{N}\otimes1_2\otimes1_2\otimes \sigma_3$.
Let us omit the SU(2) space in the above, because it is irrelevant.
A little thought tells that the Hamiltonian describes
the tangent space of Sp($N,N$)/Sp($N$)$\times$Sp($N$)
and U($N,N$)/U($N$)$\times$U($N$) for the case with
broken and unbroken time-reversal symmetry, respectively,
and accordingly, belongs to $C$II (chiral GSE) and 
$A$III (chiral GUE)
in Zirnbauer's classification \cite{Zir}.
We can now diagonalize the Hamiltonian as
$\calH=U{\rm diag}(\theta,-\theta)U^\dagger$, 
where $U$ is a unitary matrix.
The Jacobian of the change of variables from
$\calH_{ij}$ to $U_{ij}$ and $\theta_i$ is
$J={\rm const.}\Pi_{i}\theta_i^{\alpha_2}
\Pi_{i<j}(\theta_i^2-\theta_j^2)^{\beta_2}$,
where $(\alpha_2,\beta_2)=(3,4)$ and $(1,2)$ for
broken and unbroken time-reversal case, respectively.
$\beta_2$ describes the level repulsion of eigenvalues
each other, while $\alpha_2$ describes the repulsion
of each pair $\theta$ and $-\theta$ and determines the
DOS for small $\theta$ \cite{AltZir}. 
Based on this argument, we expect that
{\it 
DOS vanishes at zero energy as $E^3$ and $E$,
respectively, for the system with broken
and unbroken time-reversal symmetry.
}
Interpolating the results, we expect the behavior of DOS
as Fig. \ref{f:dos}. 
It is quite interesting to test the present conjectures
by numerical calculations.
Especially a cubic dependence of DOS on the energy
has not been known so far in disordered
$d$-wave superconductors.

The author would like to thank Y. Hatsugai, Y. Morita,
M. Shiroishi, and Mukul Laad
for valuable discussions and
J. Zittartz for his kind hospitality and encouragement.
He is supported by JSPS postdoctoral fellowship for research
abroad.



\end{multicols}


\begin{references}
\bibitem{Com}
This is valid up to four loop order \cite{Gad}.
\end{references}
\end{document}